\documentclass[twocolumn]{aastex62}

\newcommand\asec{$''$}
\newcommand\ha{H${\alpha}$}

\usepackage{natbib}

\received{April 8, 2020}
\revised{May 5, 2020}
\accepted{ May 8, 2020 }
\submitjournal{ApJ}

\shorttitle{ Distribution of Umbral Waves}
\shortauthors{Yurchyshyn et al.}

\begin{document}

\title{\sc{Spatial Distribution of Origin of Umbral Waves in a Sunspot Umbra}}

\correspondingauthor{Vasyl Yurchyshyn}
\email{vasyl.yurchyshyn@njit.edu}

\author[0000-0001-9982-2175]{Vasyl  Yurchyshyn}
\affiliation{Big Bear Solar Observatory, New Jersey Institute of Technology, \\
40386 North Shore Lane, Big Bear City, CA 92314, USA}

\author[0000-0002-0094-1762]{Ali Kilcik}
\affiliation{Faculty of Science, Department of Space Science and Technologies, Akdeniz University, 07058,
Antalya, Turkey}

\author[0000-0003-3469-236X]{Seray \c{S}ahin}
\affiliation{Department of Mathematics, Physics and Electrical Engineering, Northumbria University, Newcastle Upon Tyne, NE1 8ST, United Kingdom}
\affiliation{Faculty of Science, Department of Space Science and Technologies, Akdeniz University, 07058,
Antalya, Turkey}

\author[0000-0001-6466-4226]{Valentina Abramenko}
\affiliation{Crimean Astrophysical Observatory of Russian Academy of Science, Nauchny, Bakhchisaray, Russia}

\author[0000-0002-7358-9827]{Eun-Kyung Lim}
\affiliation{Korea Astronomy and Space Science Institute, Daedeokdae-ro 776, Yuseong-gu. Daejeon 34055, Republic of Korea}

\begin{abstract}
Umbral flashes (UFs) are emissions in the core of chromospheric lines caused by upward propagating waves steepening into shocks. UFs are followed by an expanding blue shifted umbral wave (UW) and red-shifted plasma returning to the initial state. Here we use 5~s cadence images acquired at $\pm$0.04~nm off the \ha\ line center by the Visible Imaging Spectrometer (VIS) installed on the Goode Solar Telescope (GST) to detect the origin of UFs and UWs in a sunspot with a uniform umbra free of LBs and clusters of umbral dots. The data showed that UFs do not randomly originate over the umbra. Instead, they appear to be repeatedly triggered at locations with the lowest umbral intensity and the most powerful oscillations of \ha-0.04~nm intensity. GST magnetic field measurements using Near Infra-Red Imaging Spectropolarimeter (NIRIS) also showed that the dominant location of prevalent UF origin is co-spatial associated with the strongest fields in the umbra. Interface Region Imaging Spectrograph 149.0~nm images showed that no bright UV loops were anchored in the umbra in general and near the UF patches in particular suggesting that UFs and UWs alone can not be responsible for the origin of warm coronal loops. We thus conclude that the existence of locations with prevalent origin of UFs confirms the idea that they may be driven by a sub-surface source located near the axis of a flux rope, while the presence of several UFs trigger centers may indicate the complex structure of a sunspot umbra.
\end{abstract}


\section{Introduction} \label{sec:intro}

\cite{1969SoPh....7..351B} umbral flashes (UFs) as an oscillatory type phenomenon observed in the chromosphere of sunspot umbra or a pore \cite[e.g.,][]{2013A&A...560A..84S}. They can be observed and tracked in the solar atmosphere from the photosphere to the corona over the sunspot \citep[e.g.,][]{2015A&A...580A..53L,2017ApJ...850..206S,2018A&A...618A.123S} and are thought to be driven by photospheric umbral oscillations  \citep[e.g.,][ and references therein]{1981A&A...102..147K,2000Sci...288.1396S,2002A&A...381..279T,2007A&A...463.1153T,2013A&A...552L...1B,SychR.2018a}. 

UFs are thought to be shock fronts that develop when upwardly propagating magneto-acoustic p-mode waves move across the density stratification layer in the lower atmosphere of a sunspot \citep{1970SoPh...13..323H,2010ApJ...722..888B,0004-637X-719-1-357}. The UFs are observed as blue-shifted emission due to the initial upward displacement of shocked plasma. UFs are immediately followed by a blue-shifted umbral wave and then by red-shifted plasma returning to the initial state \citep{1997ApJ...481..500C,2000SoPh..192..373B,2010ApJ...722..888B}. The temperature of the shock front may be nearly 1000~K hotter than the surrounding plasma \citep{2013A&A...556A.115D}. \cite{2019AstL...45..177Z} found the most powerful UFs to occur in the upper photosphere just below the temperature minimum. They were observed as short pulses appearing at 20~min intervals \citep[e.g.,][]{spikes,2019AstL...45..177Z}. \cite{2009ApJ...696.1683S} and \cite{2013A&A...556A.115D} reported that UFs display a fine structure with hot and cool plasma structures of sub-arcsecond scales, which was earlier predicted by \cite{2000ApJ...544.1141S} and \cite{Centeno2005} based on indirect evidence. Later, \cite{spikes} noted that the fine structure of UFs may result from spatial overlap between the bright UFs background and narrow oscillating cool umbral jets.

\cite{1972ApJ...178L..85Z} connected UFs to running penumbral waves (RPW) seen propagating from the umbra toward the edge of a sunspot and they are interpreted as upward propagating slow-mode waves guided by the magnetic field lines \citep{2013SoPh..288...73M}. Signatures of RPWs were detected in the photosphere \citep{2015A&A...580A..53L} and it has been suggested that UFs and RPWs have the same origin \citep{Tsiropoula2000,RouppevanderVoort2003,Bloomfield2007}. \cite{2017ApJ...850..206S} presented evidence that sunspot oscillations influence other wave phenomena observed higher up in the corona. The properties of these waves and oscillations can be utilized to study the inherent magnetic coupling among different layers of the solar atmosphere above sunspots. For more details on the properties of sunspot oscillation and the related phenomena please see reviews by \cite{2000SoPh..192..373B} and \cite{2006RSPTA.364..313B}.

MHD waves passing through the atmosphere of a sunspot may affect the associated magnetic field. Thus, \cite{2013A&A...556A.115D} reported 200~G variation in the penumbral fields caused by RPWs. \cite{2018ApJ...860...28H} suggested that umbral shocks may cause fluctuations of the vector magnetic field with magnitudes up to 200~G and up to 8$^\circ$, variations in the inclination and/or azimuthal angles.

\cite{1969SoPh....7..351B} and \cite{Yuan2014} concluded that UFs occur randomly anywhere over the sunspot umbra. Contrary to these, \cite{Chae2017} reported that local enhancements of 3-min oscillations were co-spatial with a light bridge (LB) and numerous umbral dots (UDs). \cite{spikes} noted that UFs tend to occur on the sunspot-center side of a LB, as well as clusters of UDs, which may indicate the existence of a compact sub-photospheric driver of sunspot oscillations. At the same time, \cite{RouppevanderVoort2003} did not find any association between UDs and UFs.

\cite{2018A&A...618A.123S} described ``background'' UFs which are weak and diffuse features without a certain shapes and localization in space and ``local'' UFs mainly located near footpoints of those magnetic loops along which 3-min propagating waves were observed. These authors thus concluded that there is a relationship between the location of the peak power of oscillations and origin of UFs.

This brief review of UFs properties indicates that there is no good understanding of the UFs origin, and in particular, the localization of their initial appearance. It may be partially because UFs are highly dynamic and rapidly evolving features and 10~s or longer cadence data does not allow us to smoothly track their evolution. Instead, we register separate, seemingly disconnected flashes that give an impression of the random nature of UFs. 

To investigate the spatial distribution of UFs and a possible spatial link between UFs and UDs, we analyzed observations of the main sunspot in NOAA AR 12384 acquired on 14 July 2015 using the Goode Solar Telescope \citep[GST,][]{2010ApJ...714L..31G, Cao2010} operating at the Big Bear Solar Observatory (BBSO). In Section 2 we describe data, in Section 3 we describe methods and present our results, and conclusions and discussions are provided in Section 4.   

\begin{figure*}[ht]
\includegraphics[width=\textwidth]{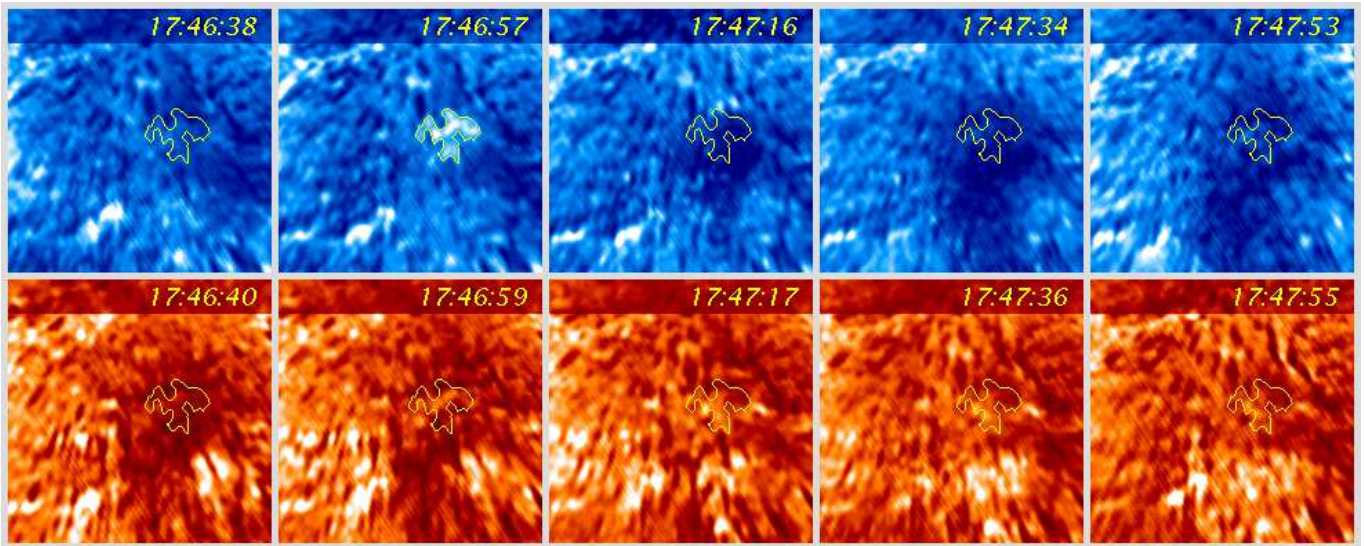}
\caption{Development of a small and bright umbral flash (UF) as seen in VIS \ha-0.04~nm (top) and \ha+0.04~nm (bottom) data. To ease the comparison, in each panel we over-plot a contour that outlines the initial shape and location of the UF as measured at 17:46:57 UT. The field of view is 10\asec$\times$10\asec. \label{ufs622}}
\end{figure*}

\section{Data} \label{sec:data}

On 14 July 2015 GST acquired a data set for a single sunspot of NOAA AR 12384 located south-west of the disk center at (175\asec,-350\asec) with the aid of the GST adaptive optics system. In this study we used three GST data sets. Photospheric TiO images taken at the 12~s cadence and the pixel scale of 0\asec.0375 were used to analyse the structure of the sunspot umbra. Next, we used data from the Visible Imaging Spectrometer (VIS) that utilizes a Fabry-P\'{e}rot interferometer with a bandpass of 0.008~nm centered at $\pm$0.04~nm off the \ha\ line center. The pixel scale was set to 0\asec.027 and images at the two spectral positions were obtained with a cadence of 5~s. All \ha\ and TiO data were speckle reconstructed using \cite{kisip_code} technique, aligned, and de-stretched to remove residual image distortion due to seeing and telescope jitter. The intensity of each image was adjusted to the average level of the data set. Finally, we also used the Fe I 1564.85~nm full-Stokes Near Infra-Red Imaging Spectropolarimeter \cite[NIRIS][]{niris,niris2}. NIRIS uses a dual Fabry–P{\'e}rot etalon that provide an 85\asec round FOV and an image scale of 0\asec.083 per pixel. The Fe I bandpass of 0.01~nm and a rotating 0.35$\lambda$ wave plate allowed us to sample 16 phase angles at each of more than 60 line positions and perform full spectra polarimetric measurements. We applied the Milne-Eddington (ME) inversion code first utilized in \cite{ChaeME} to derive the total photospheric magnetic field flux density, the inclination and azimuth angles, and the Doppler shift. This code uses a simplified model of solar atmosphere and performs very fast inversion, which is desirable when inverting a large data set.

\section{Methods and Results} \label{sec:results}

\begin{figure*}[!t]
\epsscale{1.0}
\plottwo{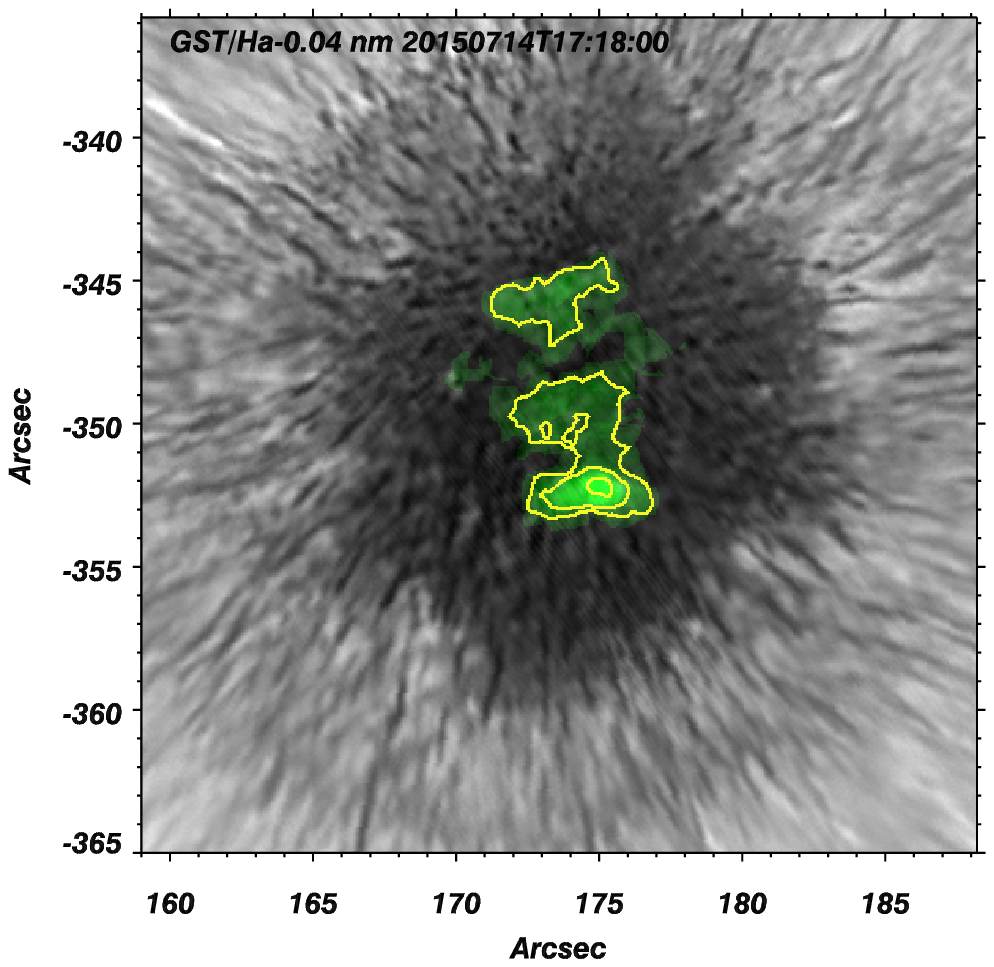}{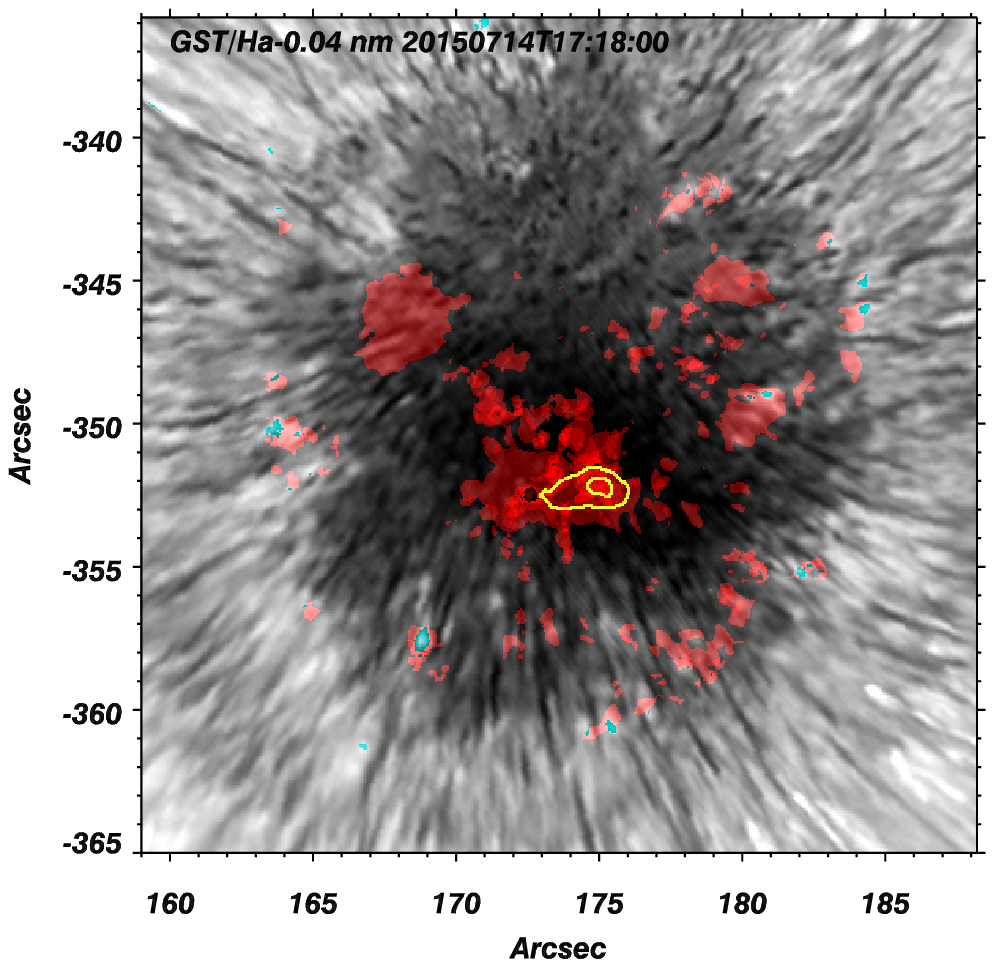}
\plottwo{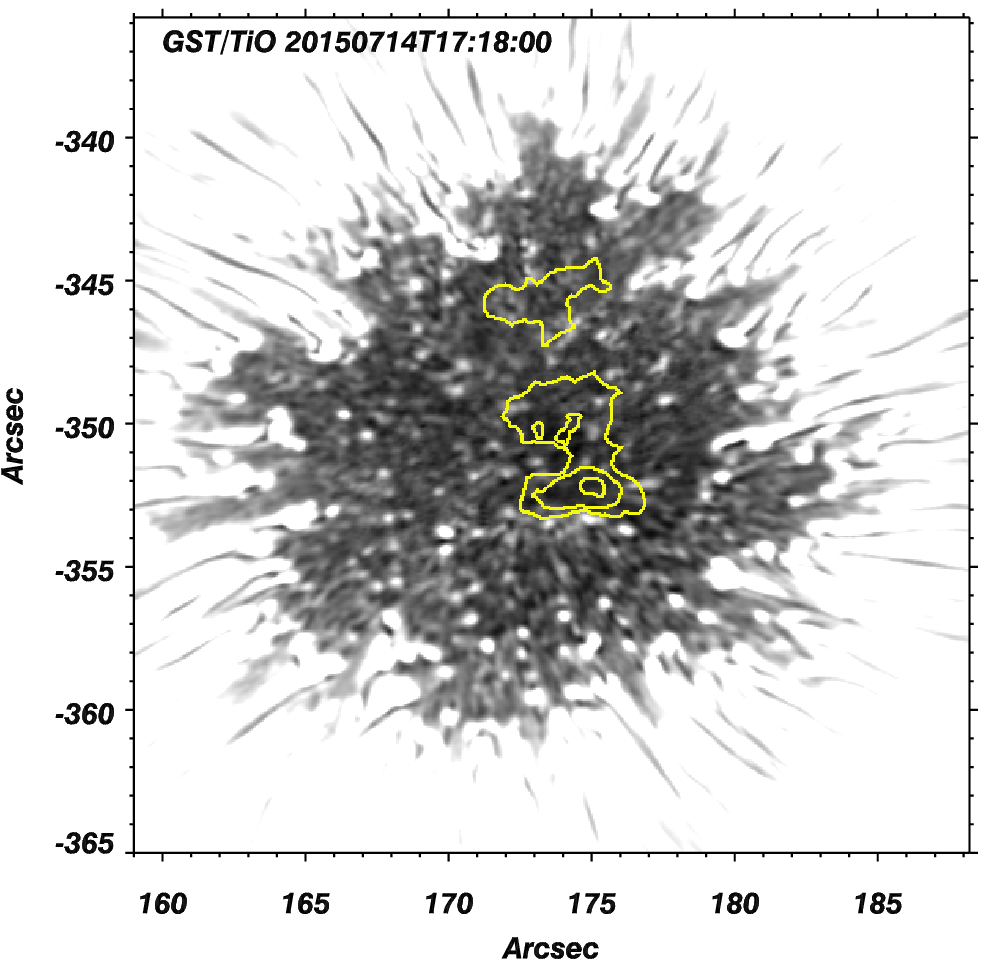}{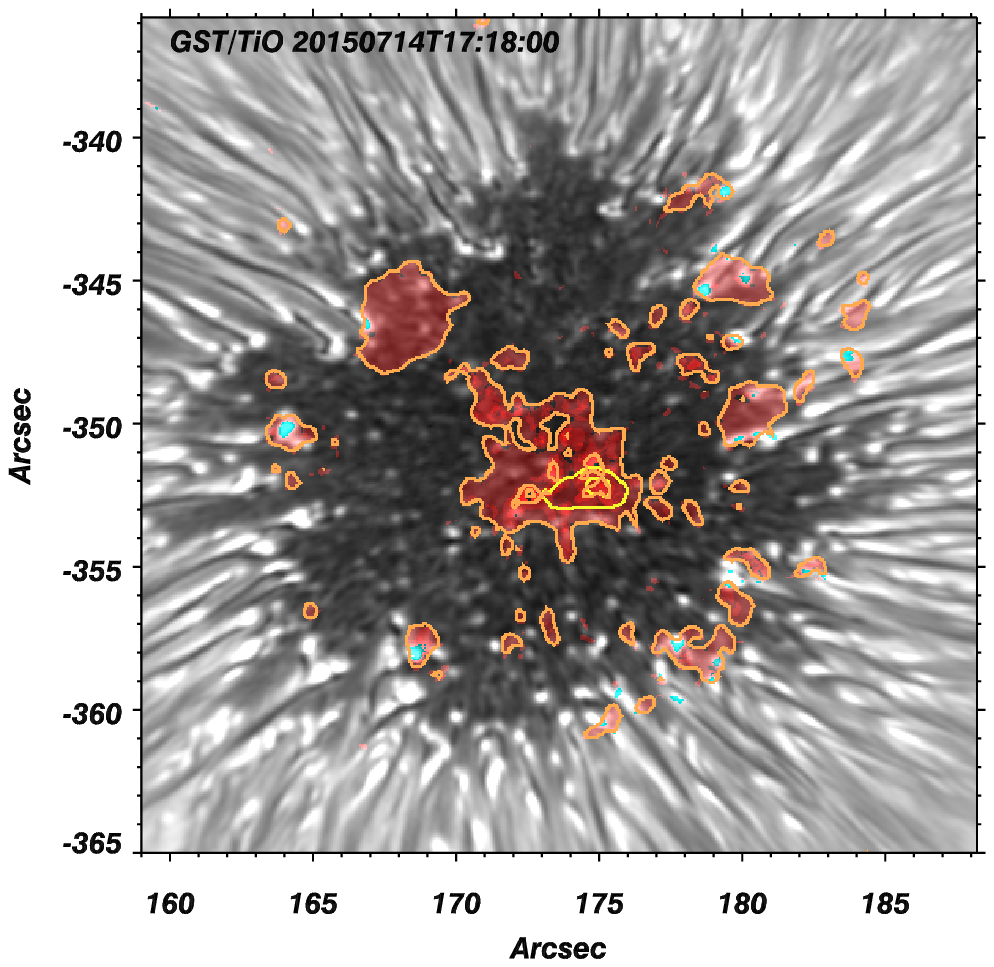}
\caption{Spatial distributions of umbral flashes (UFs) occurrence rate (green shade and yellow contours) plotted over \ha-0.04~nm (top left) and TiO (yellow contours only, bottom left) images. The TiO image was intentionally overexposed to highlight details of the umbra. The shaded area covers the pixels where UFs were detected, while the intensity indicates the number of UFs that occurred at a given pixel. The yellow contours are plotted at 5, 10, 20 events per pixel levels. Similarly, the red shaded areas in the right panels indicate the spatial distribution of the occurrence rate of strongest (darkest) umbral waves plotted over \ha-0.04~nm (top right) and TiO (bottom right) images. The over yellow plotted contours are the same as in the left panels. \label{ufs}}
\end{figure*}

Figure \ref{ufs622} shows a small UF developing over a 10\asec$\times$10\asec fragment of the sunspot umbra as it is seen in \ha$\pm$0.04~nm VIS images. The sub-frame is centered at 175\asec,-350\asec (see Figure \ref{ufs}). The UF appeared at 17:46:57~UT at the end of the downflow branch of the oscillation cycle (see 17:46:40 and 17:46:59~UT \ha+0.04~nm panels) and it was very rapidly replaced by strong blue-shifted upflows (see 17:47:16~UT \ha-0.04~nm panel, where the contour indicates the position and shape of the initial UF), which gave rise to a dark umbral wave (UW) that further evolved into an RPW. The expanding UWs was asymmetrical and no detectable signature of UFs riding at the front of the UW \cite[e.g.,][]{Yuan2014} were observed in this particular UF event. In this study we will refer to such events as the origin of UWs. 

Our goal was to explore the origin of sunspot oscillations by analysing the spatial distribution of the UW origin. We employed three various techniques. First, using the available 90-min long data set consisting of 688 red and 688 blue wing \ha\ images, and based on visual inspection we manually outlined those areas where the first signatures of a new UF appeared (see Figure \ref{ufs622}). The diffuse UFs riding at the front of expanding UWs were not taken into account. After inspecting all images in the data set we constructed a map of spatial distribution of the UF occurrence rate where each pixel value indicated the number of times a given pixel was associated with the origin of UFs. Second, using the same data set, in each image we automatically determined areas occupied by the most intense (darkest) UWs by using low pass image filtering and thresholding. Similarly, a map of spatial distribution of the UW occurrence rate was constructed, where each pixel value indicated the number of times the given pixel was covered by an UW. Finally, following \cite{Chae2017} and \cite{2018ApJ...852...15P} we used a series of \ha-0.04~nm images to calculate a power spectrum of intensity oscillations at each pixel of the image and then integrated the power within the 30~s--4~min time interval. 

The results are  shown in Figure \ref{ufs}, where the top panels are \ha-0.04~nm images shaded with the UF (green, left) and UW (red, right) occurrence rate maps. The more intense shade of green indicate areas where UF frequently originated, while the more intense shade of red marks areas where the most intense dark UW were most frequently detected. We note the following. First, UFs do not randomly occur over the umbra. Instead, there are two distinct locations: i) dominant lower UF patch (centered at 175\asec,-351\asec, peak at 23 events/pixel) and ii) weaker upper UF patch (176\asec,-346\asec, 10 events/pixel), where UFs appeared to originate more frequently. Other extended areas counted less than 7 events per pixel. By comparing the two top panels we also conclude that the dominant lower UFs patch (yellow contours in the upper right panel) spatially coincides with the location where UW occur most frequently. In the lower left panel we show an overexposed co-temporal photospheric TiO image with the overplotted UF occurrence contours. In general, the umbra of this sunspot was mainly uniform and the UF and UW patches overlap with both dark umbral areas and UDs. We would like to point out, however, that there is a relatively small ($\approx$2\asec squared) area  centered at 174\asec,-352\asec.5 void of UDs located right under the lower green UF and red UW patches. It is also worth noting the lower edge of that patch is bordered by a distinct chain of UDs running from (172\asec,-353\asec.5) to (177\asec,-353\asec.5). Although this UD chain can not be classified as a sunspot light bridge, it is interesting that \cite{spikes} noted that UFs tend to occur on the sunspot-center side of a LB, as well as clusters of UDs, while \cite{Chae2017} found enhanced 3-min oscillations to be co-spatial with a LB and numerous UDs.

\begin{figure*}[!t]
\epsscale{1.1}
\plottwo{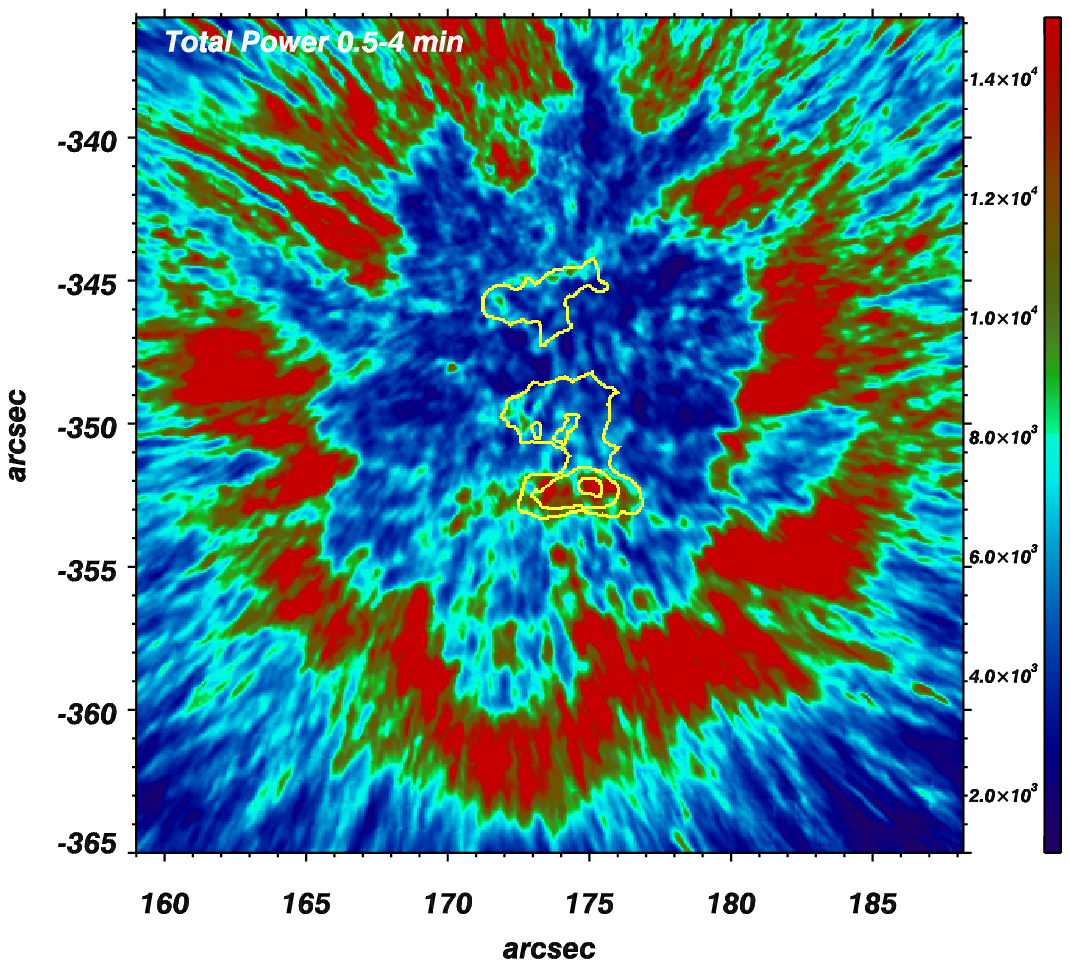}{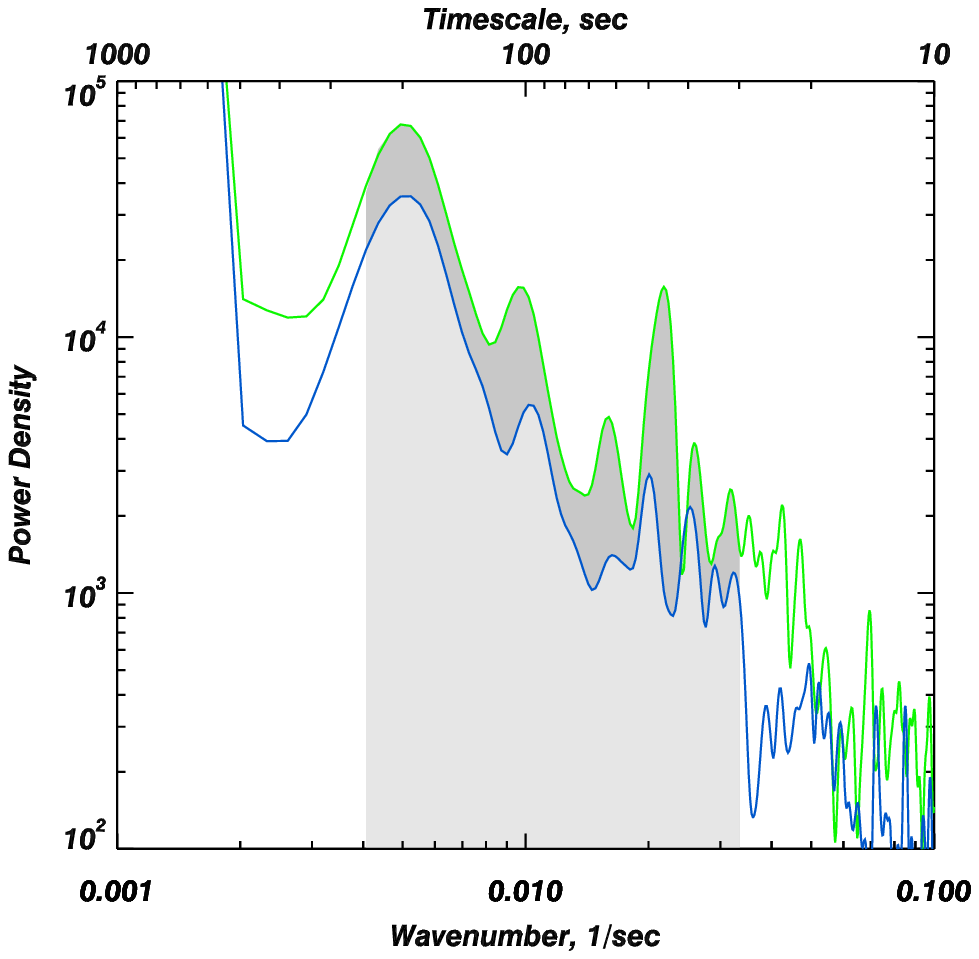}
\caption{Left: Total power of \ha-0.04~nm intensity oscillations integrated between 30~s and 4~min frequencies. The red shaded area represent locations with highest power of intensity oscillations. The contours are the same as in Fig. 2 and outline the area of UFs production. Right: The green curve represents a power spectrum calculated for the location with the highest UFs occurrence rate (red shaded patch at 175\asec,-353\asec\ within the smallest contour), while the blue curve is a power spectrum for an adjacent dark blue shaded area at (177\asec,-351\asec), where UFs never initiated. The shaded areas indicate the power integration interval. \label{ps}}
\end{figure*}

In Figure \ref{ps} (left) we show the 0.5-4~min power map with the red/blue colors indicating high/low oscillating power. The overplotted UF occurrence contours suggest a possible connection between UFs and locations of enhanced oscillations (as compared to the rest of the umbra), which generally shows lack of power at these timescales (extended blue shaded area). In the right panel of Figure \ref{ps} we show two power spectra determined at i) location of the most frequent UF occurrence (inside the smallest UF contour) and ii) immediately next to it (177\asec,-351\asec) inside an area with a weak oscillating power. While the both spectra have similar general appearance the total power of the associated oscillations differs from one case to another.

\begin{figure}
\epsscale{2.1} 
\plottwo{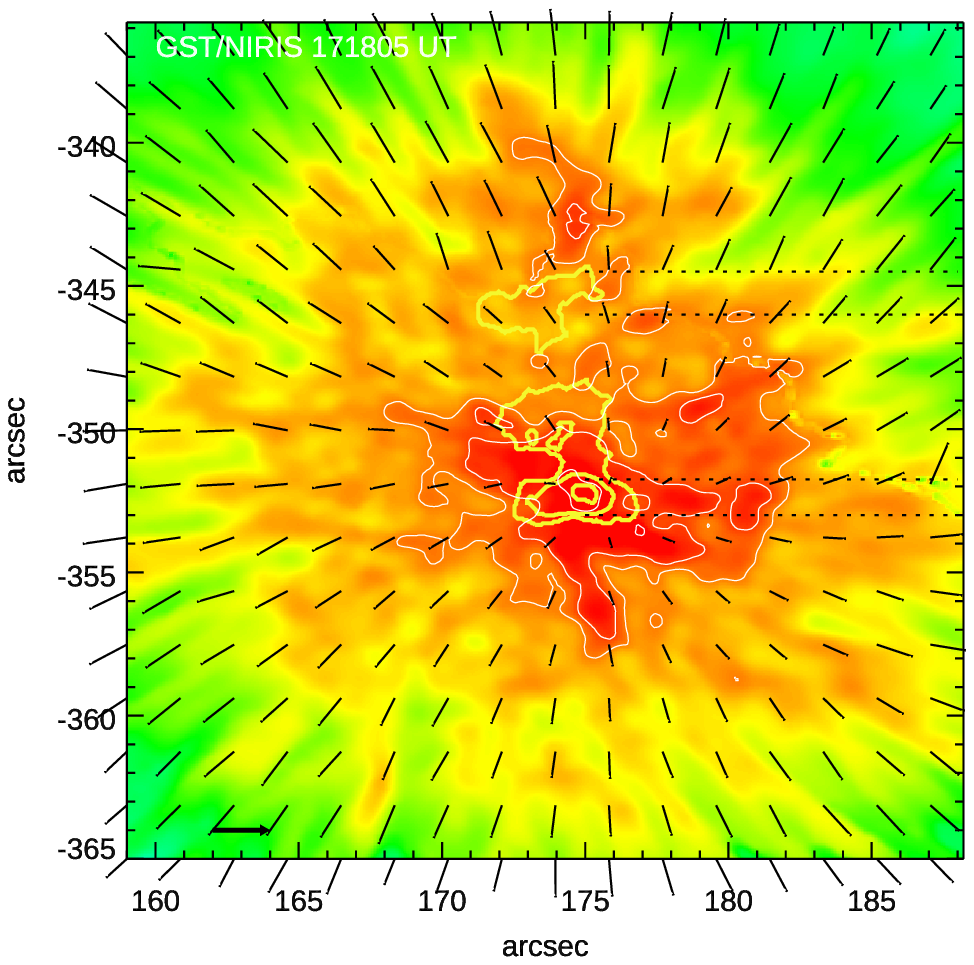}{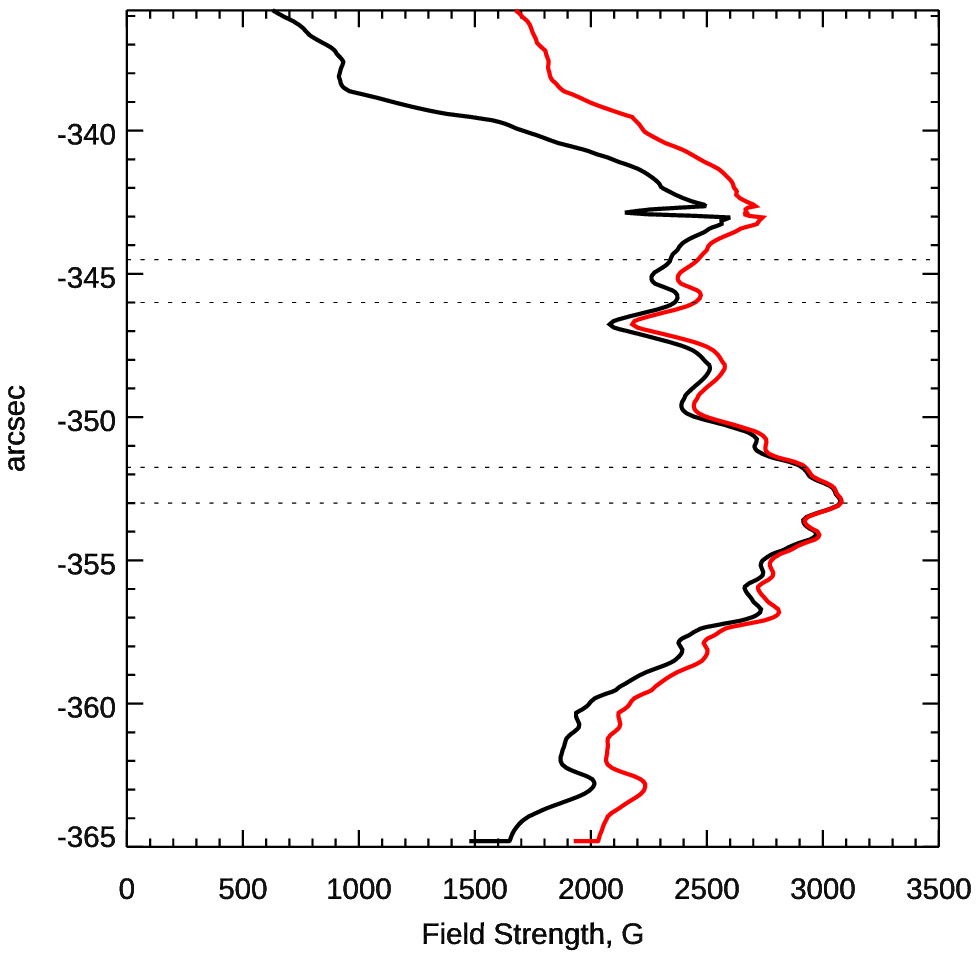}
\caption{Left panel: NIRIS transverse fields (line segments) plotted over the corresponding magnitude of the total magnetic field scaled between 0 (blue) and 2700~G (red). The white contours are drawn at 2500, 2700, and 3000~G levels. The yellow contours are the UFs occurrence contours the same as in Fig. \ref{ufs}. The horizontal arrow at the bottom corresponds to 2000~G transverse fields.  Right panel: Line-of-sight component (black) and total field (red) along a vertical cut at $x=175$\asec\ in the left panel (not shown). The two pairs of horizontal dotted lines in each panel indicate the position of two UFs centers.} \label{niris}
\end{figure}

To explore the spatial relationship between the occurrence of UFs and the sunspot magnetic field, in Figure \ref{niris} we show the total photospheric magnetic field strength ($B_{tot}$, left panel, background) and the corresponding transverse field (line segments), corrected for the projection effect. The upper UF patch is situated in an area of mostly vertical 2400~G fields next to a small local peak of 2700~G. The lower UFs patch is entirely located in a vast area of 2700~G fields with the most intense UF production area coinciding with a 3000~G peak of the umbral field (Figure \ref{niris}, right panel). It appears that both UF patches are located on the sunspot-center side of the enhanced magnetic fields (see dashed lines in the right panel).

\begin{figure}
\epsscale{1.1}
\plotone{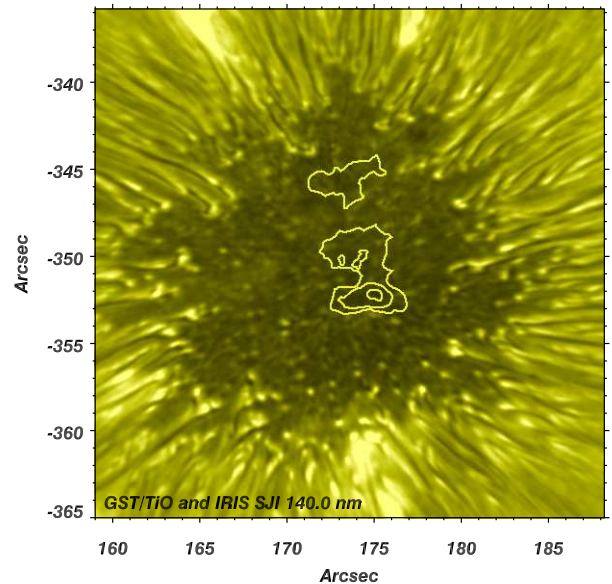}
\caption{A composite image showing photospheric GST/TiO image (umbra and penumbra) and bright 140.0~nm patches (160\asec,-357\asec; 175\asec,-362\asec; 160\asec,-337\asec; and 177\asec,-337\asec) from the corresponding IRIS slit-jaw image overplotted with the UFs contours. \label{iris}}
\end{figure}

Finally, in Figure \ref{iris} we show a composite GST/TiO (umbra and penumbra) and IRIS 149.~nm (brightness enhancements) image overplotted with the UF contours. The four bright 140.0~nm patches located at (160\asec,-357\asec; 175\asec,-362\asec; 160\asec,-337\asec; and 177\asec,-337\asec) are footpoints of bright UV loops systems rooted in the inner penumbra and no loops are seen anchored in the umbra and near the UF patches in particular.  


\section{Summary and Discussion}

Analysing data for one sunspot \cite{Chae2017} and \cite{2019ApJ...879...67C} found that power of 3-min oscillations is enhanced in vicinity of a LB and around UDs. \cite{Chae2017} also suggested that magnetoconvection in the LB and UDs may drive upwardly propagating slow magneto-acoustic waves in the photosphere \citep[also see e.g.,][]{2009A&A...505..791S, 2012SoPh..279..427K,2016ApJ...817..117S}. \cite{spikes} found that UFs tend to occur on the sunspot-center side of LBs and clusters of UDs, which may indicate the existence of a compact sub-photospheric driver of sunspot oscillations. To further explore a possible role of LBs, we intentionally chose for this study a sunspot with a relatively uniform umbra (no multiple umbral cores) free of LBs and extended clusters of bright and large UDs. Our data showed that even in the absence of LBs, UFs do not randomly originate over the umbra of a sunspot. Instead, the detected UFs/UWs appeared to be triggered more frequently at locations associated with strong magnetic fields. The location of the prevalent UF origin (the lower UF patch) is co-spatial with the darkest umbra and the most powerful oscillations of \ha-0.04~nm intensity. We should note that the dominant UF production region was located on the sunspot-center side of a narrow chain of UDs, which agrees with \cite{spikes}. However, there were other UD clusters/chains of similar intensity not associated with UFs origin suggesting that the presence of LBs and UDs may not be sufficient condition for UF/UW generation. This inference seem to agree with that of \cite{RouppevanderVoort2003} who concluded that UDs do not play an ``obvious role'' in UF generation. Instead, \cite{1993A&A...267..275A} emphasized the effect of the magnetic field strength and topology on the intensity of 3~min oscillations, while \cite{2016ApJ...817..117S} suggested that the propagation of UWs may be affected by the twist of the umbral magnetic field. \cite{10.1093/pasj/63.3.575} detected the origin of 19 umbral wave events in a sunspot with LBs. Although at the first sight the UW sources there appear to be scattered over the umbra \citep[see ``+'' symbols in Fig. 1 of][]{10.1093/pasj/63.3.575}, about half of them are located within a small 5\asec$\times$5\asec area, which is only a fraction of the entire umbra under study and is comparable to the area of the lower UF patch discussed in this study. Thus \cite{10.1093/pasj/63.3.575} data agrees with the conclusions drawn in the current study.

At the same time our conclusions seem to be in direct contradiction with other studies reporting a random occurrence of UFs over the sunspot umbra \cite[e.g.,][]{1969SoPh....7..351B,RouppevanderVoort2003,Yuan2014}.  Also, \cite{2018ApJ...852...15P} reported that most of the  umbral oscillations initially emerge either at or close to the umbra-penumbra boundary and also are a part of the preceding RPWs. Authors suggested that nearly all UWs are connected to the preceding RPWs. There are two reasons that together may account for the discrepancy. First, we did not track each UF brightening detected in the umbra. Instead, we were only interested in the locations where an UF and UW were triggered, discarding such events as dynamic UFs riding at the front of UWs. We thus find that UF/UW originate within the umbra and no significant events were detected at the umbral boundary. We also would like to emphasize that temporal resolution is critical for proper understanding of the dynamics of UFs and UWs. Second possible reason is that we analysed a simple sunspot with a symmetrical umbra free of LBs and large clusters of UDs. Earlier studies analysed sunspots with multiple umbral cores separated by various LBs (narrow and wide). Each of these core could be source of oscillations together creating a complex and very dynamic wave pattern. Thus, \cite{1993A&A...267..275A} reported absence of correlation between different umbral cores, while \cite{Yuan2014} found umbral velocity oscillations at both sides of a LB to be in phase, although the corresponding UFs did not show similar tendency. \cite{2016ApJ...817..117S,Su2016} suggested that UWs triggered at different umbral cores could interfere at LBs, while \cite{2016ApJ...821L..30K} reported enhancement of 3-min oscillations in the sunspot umbra triggered by impulsive downflows \citep[also see][]{2015ApJ...808..118C}. \cite{2019A&A...621A..43F} studied spiral wave patterns observed in sunspot umbrae, while \cite{Kang2019} modeled them as superposition of two different azimuthal modes of slow magnetoacoustic waves driven from below the surface in an untwisted and non-rotating magnetic cylinder. All of these further complicate the resulting oscillating pattern in sunspots with multiple umbral cores.

\cite{SychR.2018a} distinguished background and local UFs. The background UFs were defined as weak and diffuse brightenings that ride the UW fronts, the local UFs sources are those that appear in close proximity to the footpoints of umbral loops with strong propagating of 3-min oscillations. These authors suggested that the local UF sources are related to footpoints of those umbra anchored loops with enhanced 3-min oscillations. The UF events that we analysed can be classified per \cite{SychR.2018a}'s definition as local UFs. According to AIA and IRIS data, no UV loops were observed rooted in this uniform umbra free of LBs, which indicates that there is no simple relationship between UFs and the oscillating loops. Some additional mechanisms must be invoked to fill umbra anchored loops with plasma dense enough for them to become visible in the UV images. Earlier studies \cite[e.g.,][]{Tian_2014,2015ApJ...798..136Y} pointed out a connection between the footpoints of coronal loops, bright transition region umbra and magnetoconvection features such as LBs and clusters of UDs. At the same time, \cite{Sahin2019} reported that warm coronal loops rooted in the umbra may be heated due to small-scale reconnection process occurring at the remote, non-sunspot footpoint of the loop. 

Finally, we note that there were two patches of UF and UW origin and the most intense one is co-spatial with the maximum of the umbral field, while the weaker upper UF patch was located in the vicinity of a much smaller but distinct maximum. These suggest that this seemingly simple sunspot may have been composed of two umbral cores of different intensity each of them having its own sub-photospheric driver of oscillations. \cite{0004-637X-719-1-357} used 3D MHD simulations to show that a source of oscillations located near the axis of a sunspot may create a pattern of shocks (UFs and UWs) propagating away from the sunspot center, which is very similar to the dynamics of UFs reported here and in \cite{2015ApJ...798..136Y}. Similar conclusion was reached in \cite{2016ApJ...830L..17Z}, who placed a wave source of at 5~Mm beneath the photosphere. While the nature of this wave source seems to be p-mode waves the force behind it remains unclear. One possibility is that the driver results from interaction of quiet Sun p-mode waves that enter a sunspot from beneath with the magnetized sunspot plasma. Another one suggests that the oscillations are exited by an internal driver. Thus, \cite{ChoKH2019} detected several oscillation centers above rapidly evolving UDs, which they regarded as evidence that magnetoconvection associated with UDs inside sunspots can drive the 3-min umbral oscillations.

\acknowledgments

We thank the anonymous referee for helpful comments on the manuscript. BBSO operation is supported by NJIT and US NSF AGS-1821294 grants. GST operation is partly supported by the Korea Astronomy and Space Science Institute (KASI), Seoul National University, the Key Laboratory of Solar Activities of Chinese Academy of Sciences (CAS), and the Operation, Maintenance and Upgrading Fund of CAS for Astronomical Telescopes and Facility Instruments. V.Y. acknowledges support from NSF AST-1614457, AFOSR FA9550-19-1-0040, NASA 80NSSC17K0016, 80NSSC19K0257, and 80NSSC20K0025 grants. This study was partially supported by Scientific and Technical Council of Turkey Project 117F145 and by basic research funding from Korea Astronomy and Space Science Institute (KASI).

\vspace{5mm}
\facility{GST}



\end{document}